 \documentclass[twocolumn]{aastex6}

\begin{document}


\title{Boxy/Peanut/X-shape bulges: steep inner rotation curve leads to \\ barlens face-on morphology}


\author{H. Salo\altaffilmark{1} and E. Laurikainen \altaffilmark{1}}
\affil{Astronomy Research Unit, University of Oulu, FI-90014 Finland\\
\email{heikki.salo@oulu.fi}
}
  \date{Accepted to ApJ, November 8th, 2016}  


\begin{abstract}
We use stellar dynamical bulge/disk/halo simulations to study whether
barlenses (lens-like structures embedded in the narrow bar
component) are just the face-on counterparts of Boxy/Peanut/X-shapes (B/P/X) seen in edge-on bars, or if some additional
physical parameter affects that morphology.  A range of
bulge-to-disk mass and size ratios are explored: our nominal parameters ($B/D=0.08$, $r_{\rm eff}/h_r=0.07$, { disk comprising 2/3 of total force at
$2.2h_r$}) correspond to typical  MW mass
galaxies.  In all models a bar with pronounced B/P/X forms in a few
Gyrs, visible in edge-on view. However, the pure barlens
morphology forms only in models with sufficiently steep inner rotation
curves, $dV_{cir}/dr\gtrsim5V_{max}/h_r$, achieved when
including a small classical bulge with $B/D\gtrsim0.02$ and $r_{\rm eff}/h_r\lesssim0.1$. For shallower slopes the central structure
still resembles a barlens, but shows a clear X-signature even in low
inclinations.  Similar result holds for bulgeless simulations,
  where the central slope is modified by changing the halo
  concentration. 
The predicted sensitivity on inner rotation curve is consistent with the slopes
  estimated from gravitational potentials
calculated from the 3.6$\mu$m images, for the observed barlens and
X-shape galaxies in the Spitzer Survey of Stellar Structure in
Galaxies (S$^4$G). For inclinations $<60^\circ$ the galaxies with
barlenses have on average twice steeper inner rotation curves than galaxies
with X-shapes: the limiting slope is $\sim250$km/s/kpc. 
Among barred galaxies, those
with barlenses have both the strongest bars and the largest relative
excess of inner surface density, both in barlens region
($\lesssim0.5h_r$) and near the center ($\lesssim0.1h_r$);
this provides evidence for bar-driven secular evolution in galaxies.
\end{abstract}





\section{Introduction} \label{sec:intro}

Majority of strong bars in massive early type disk galaxies exhibit a
{barlens} (bl) morphology: a central lens-like structure embedded
in a narrow bar component, identified as a distinct morphological
feature in \citet{laurikainen2011}.  Recent observational and
theoretical studies support the conjecture \citep{laurikainen2007}
that such round inner structures are in fact the face-on counterparts
of the vertically extended {Boxy/Peanut/X-shape} (B/P/X)
structures commonly seen in barred edge-on galaxies
\citep[{e.g }][]{lutticke2000, bureau2006}. Namely, the stellar masses
and the axial ratio distributions of their host galaxies are
consistent with B/P/X and bl galaxies forming a single population
\citep{laurikainen2014,laurikainen2016a}. Simulation models also show that it is
possible to exhibit a B/P/X morphology in edge-on view while the
face-on morphology is essentially circular \citep{athanassoula2015,athanassoula2016a}.
Additional support for the B/P/X/bl connection is provided by the very
similar colors of the barlens and narrow bar components
\citep{herrera2016}.  This connection of barlenses to vertically
extended inner portions of bars indicates that they should not be
confused with merger-related classical bulges, or with disky
'pseudo-bulges' ({\em i.e} inner disks).

The best examples of pronounced barlens morphology in simulations are
those of \citet{athanassoula2013}, which simulations were analyzed in
detail in the above mentioned study of \citet{athanassoula2015}.
These idealized isolated galaxy N-body + SPH simulations followed the
evolution of a stellar+gaseous disk embedded in a tri-axial halo, and
included recipes for converting gas into stars. The closest
resemblance to observed barlens morphologies was attained with initial
gas fractions 20\%-80\% \citep{athanassoula2015}. Recently, barlens
morphology was also reported in simulations where disk galaxies were
formed as a result of gas-rich mergers \citep{athanassoula2016}.
However, in most corresponding high-resolution simulations, either
with or without gas \citep[e.g.][]{minchev2012, saha2012,
  dimatteo2013}, the formed bars have elongated shapes without round
inner barlens components.  Also, a close inspection of
\citet{athanassoula2015} models indicates that even there many of the
simulated images (see their Fig. 2) contain a slight trace of X-shape
morphology in the face-on view.  Such X-shape signatures are very rare in the
observed nearly face-on barlens galaxies \citep{laurikainen2016b},
suggesting that the orbital families forming the simulated central
structures are not exactly the same as in most observed galaxies.

In this study we perform simple stellar dynamical
bulge/disk/halo simulations to address the conditions for obtaining
the observed type  pure barlens morphology. We vary the
parameters influencing the slope of the inner rotation curve, and show
how this affects the face-on morphology of the vertically extended
inner bar region. The S$^4$G survey \citep[Spitzer Survey of Stellar
  Structure in Galaxies;][]{sheth2010} is used as a guide for choosing
the simulations parameters, and also as a comparison sample to check
the predictions of the simulations.

\section{Simulations} \label{sec:simu}

We perform stellar dynamical N-body simulations with GADGET-2
\citep{springel2005}. The self-consistent initial galaxy models are
constructed with the GalactICS software \citep{kuijken1995}, and they
consist of an exponential disk, and a lowered Evans-model (truncated
log-potential with a core) for the spherical halo.  Compared to the
N-body-SPH simulations in \citet{athanassoula2013}, our simulations
contain no gas or star-formation; we also include a small classical
bulge (King-model) in the initial conditions, whereas in their study
central components could form during the simulation via gas inflow and
star formation.

\begin{figure}[ht!]
\figurenum{1}
\epsscale{1.25}
\plotone{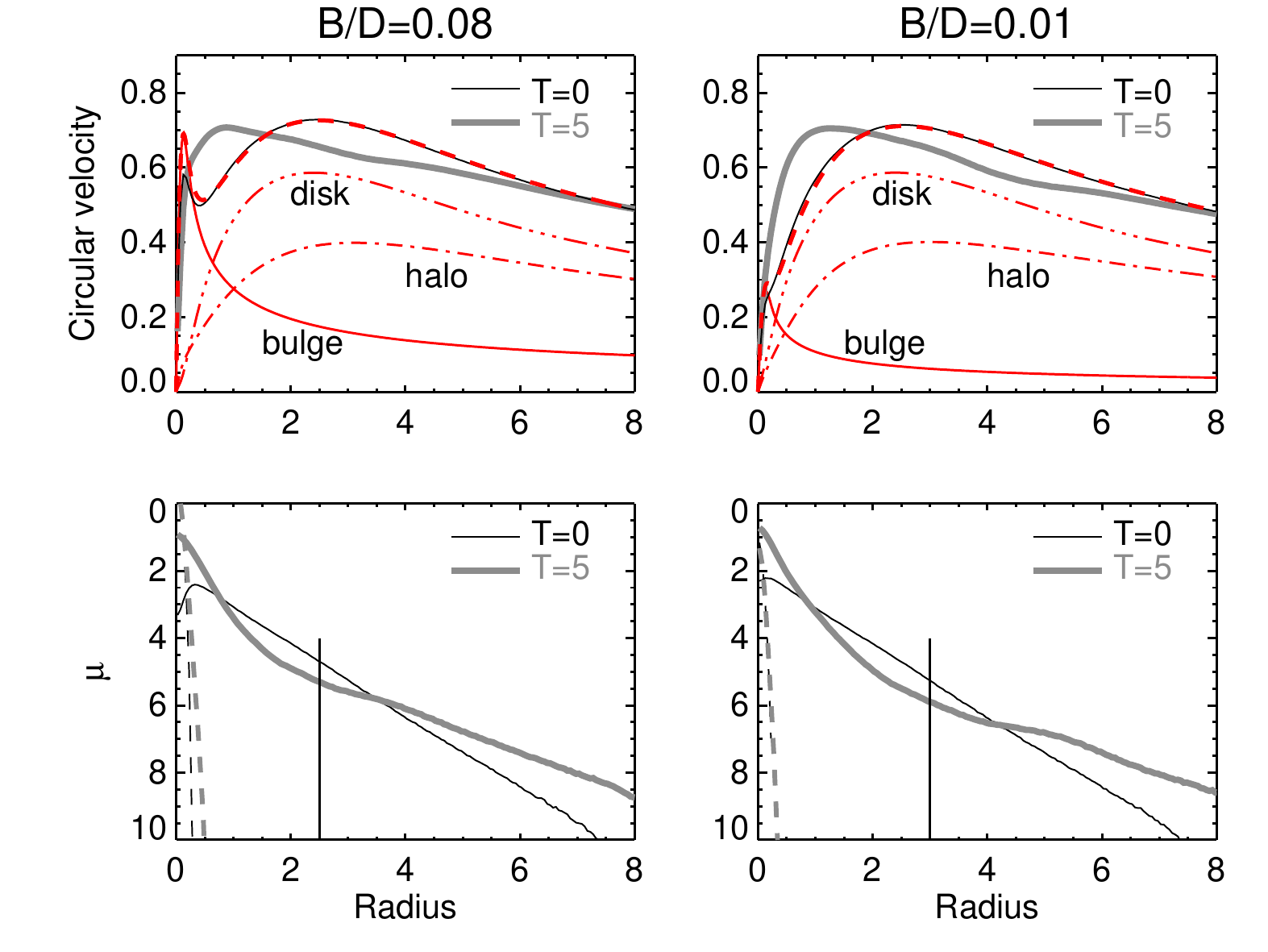}
\caption{The simulation models. The upper row compares the rotation
  curves in two simulations with bulge-to-disk mass ratio $B/D=0.08$
  (left) and $B/D=0.01$ (right). The red curves indicate the initial
  total rotation curve model from GalactICS, together with separately
  shown bulge, disk, and halo contributions. The black curves are the
  actual total circular velocity curve corresponding to the softened forces
  calculated by GADGET: thin curve refers to initial state and thick
  curve is after 5 Gyrs. The radius is indicated in units of initial disk
  scale length; one velocity unit equals $\sim 300$ km/sec when using
  the scaling to physical units discussed in the text.  The lower
  frames show the azimuthally averaged surface densities (in magnitude
  units, with arbitrary zeropoint) of the disk (solid lines) and bulge
  (dashed lines). Again, thin lines refer to the initial model, while
  thick lines are after 5 Gyrs.
\label{fig_inival}
}
\end{figure}

\begin{figure}[ht!]
\figurenum{2}
\epsscale{1.25}
\plotone{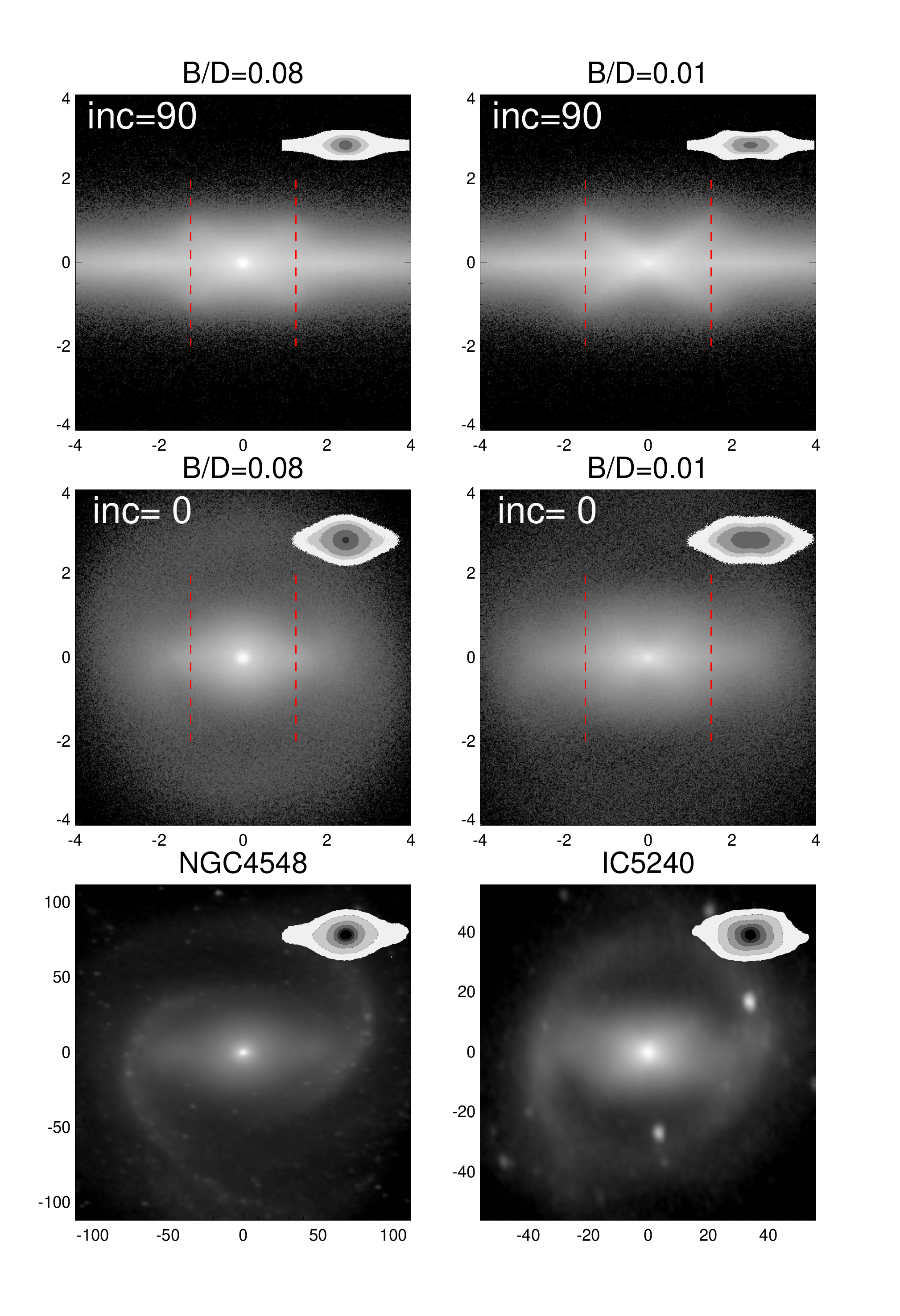}
\caption{ Two simulation models with a small classical bulge:
  $B/D=0.08$ {\em (left)} and $B/D=0.01$ {\em (right)}. Both lead to a
  B/P/X shape structure when seen edge-on ($i=90^\circ$, upper row).
  In face-on view ($i=0^\circ$, middle) the $B/D=0.08$ simulation has
  a morphology very similar to observed barlens galaxies, while the
  model with $B/D=0.01$ shows a trace of X-shape signature. The
  difference is particularly clear in the small insert figures,
  showing the isophotal contours of the central regions. The lowermost
  row shows two deprojected images from the S$^4$G-survey, NGC 4548
  ($i=39^\circ$, with a barlens), and IC 5240 ($i=44^\circ$, with an
  X-shape). In the upper two rows the axis units are in 
    simulation units (initial scale lengths), in the lowermost row in
  arcseconds.
\label{fig_bd}
}
\end{figure}

Our nominal model parameters are based on the analysis of the 3.6
$\mu$m S$^4$G images: to mimic a typical Milky Way (MW) galaxy (with
stellar mass $M^* \sim 5\cdot 10^{10} M_\sun$) we use a bulge-to-disk
mass ratio $B/D=0.08$ and a bulge effective radius $r_{\rm
  eff}/h_r=0.07$, where $h_r$ denotes the scale length of the disk:
these were typical bulge values obtained for Hubble types $T=3$ in the
S$^4$G  pipeline bulge/disk/bar decompositions \citep{salo2015}. More detailed
decomposition, using a separate component for a bar and a barlens
\citep{laurikainen2014}, indicated similar small $B/D$ values even for
galaxies with $-2 \le T \le 0$.  Our halo model is chosen so that at
{$2.2h_r$ the disk accounts for 65\% of the total radial force}. This
corresponds to the typical value estimated for a MW mass galaxy in
\cite{diaz2016a}, obtained by combining the gravitational field
calculated from the 3.6 $\mu$m images using the NIRQB-code
\citep{salo1999, laurikainen2002}, with the rotation amplitudes
obtained from the HI-kinematics \citep{courtois2009}.  The initial
{Toomre parameter  $Q_T \approx 1.35$ at $2.2h_r$}, and the vertical
thickness of the disk has a constant scale height $z_o=0.2h_r$, which
corresponds to { ratio of vertical to radial velocity dispersion
$\sigma_z/\sigma_r \approx 0.6$ at  $2.2h_r$.}  

The simulation units are
fixed by setting the gravitational constant, the mass, and the initial
scale length of the disk to unity. To convert to physical units,
  we identify $m_D=5 \cdot 10^{10} M_\sun$ and $h_r=2.5$ kpc, which
  corresponds to maximum disk contribution to circular velocity $\sim
  170$ km/sec (total circular velocity $\sim 215$ km/sec; see
  Fig. 1).  Our standard value for gravity softening is
$\epsilon=0.01 h_r$, and the time step is $\sim 10^{-2}$ time units;
in physical units one time unit corresponds to $\sim 10$ Myrs. 
  The disk is realized with $10^6$ particles, each with mass of $5
  \cdot 10^4 M_\sun$. The halo and bulge particles have the same mass
  as the disk particles.  

The rotation curve and the disc \& bulge surface density profiles of our
nominal simulation model with $B/D=0.08$ are displayed in Fig. 1, both
for the initial state and for $T=5$ Gyrs, after a bar has formed. An
otherwise similar model except that  $B/D=0.01$ is also shown.  As
expected \citep[see e.g ][]{debattista2006}, the formation of the bar
modifies the disk density profile, which becomes more centrally
concentrated while the scale length of the outer disk
increases. This also modifies the rotation curve, thought the
central slopes are little affected. In all the following, the discussed
rotation curves correspond to those after the bar has already formed.

Figure 2 compares the morphology in simulations with two
different bulge-to-disk mass ratios, $B/D =0.08$ (left) and $B/D=0.01$
(right). Both simulations form a bar during $T=1-2$ Gyrs, and while
the bar starts to form, also the vertical extent of the orbits in the
inner part of the bar increases.  At $T=5$ Gyrs (shown in the figure)
this has resulted in both cases to a pronounced B/P/X edge-on
morphology.  The vertical evolution lacks a rapid buckling phase often
reported in the literature \citep{raha1991}, in which sense it is more
reminiscent to the resonance heating models \citep{quillen2014} where
the X is associated to disk stars heated by the 2:1 vertical
resonance \citep{combes1990, pfenniger1991, patsis2002}.  Most
importantly for the current study, the face-on morphology for
$B/D=0.08$ is very similar to that in massive barred galaxies, with a
dominant round barlens structure and a weak thin bar component
(compare to NGC 4548 in the lowermost row; see \cite{laurikainen2016b}
for classification of observed barlens categories): this structure
survived to the end of the simulation ($12$ Gyrs).  In face-on view
the size of the barlens, and that of X-feature in edge-on view, is
about 1/2 of bar size, similar to what was found in
\cite{athanassoula2015}.  In the run with $B/D=0.01$ the face-on
morphology also resembles a barlens, but with a clear waist-like
narrowing on the bar minor axis.  Most of the simulations in
\cite{athanassoula2015}, as well as that in Fig. 1 in
\cite{laurikainen2014} resemble more the simulation with $B/D=0.01$
than that with $B/D=0.08$.  Although such traces of X-like morphology
are not often seen in observed nearly face-on galaxies, one example is
IC 5240, with an inclination of $i=44^\circ$.

\begin{figure}[ht!]
\figurenum{3}
\epsscale{1.2}
\plotone{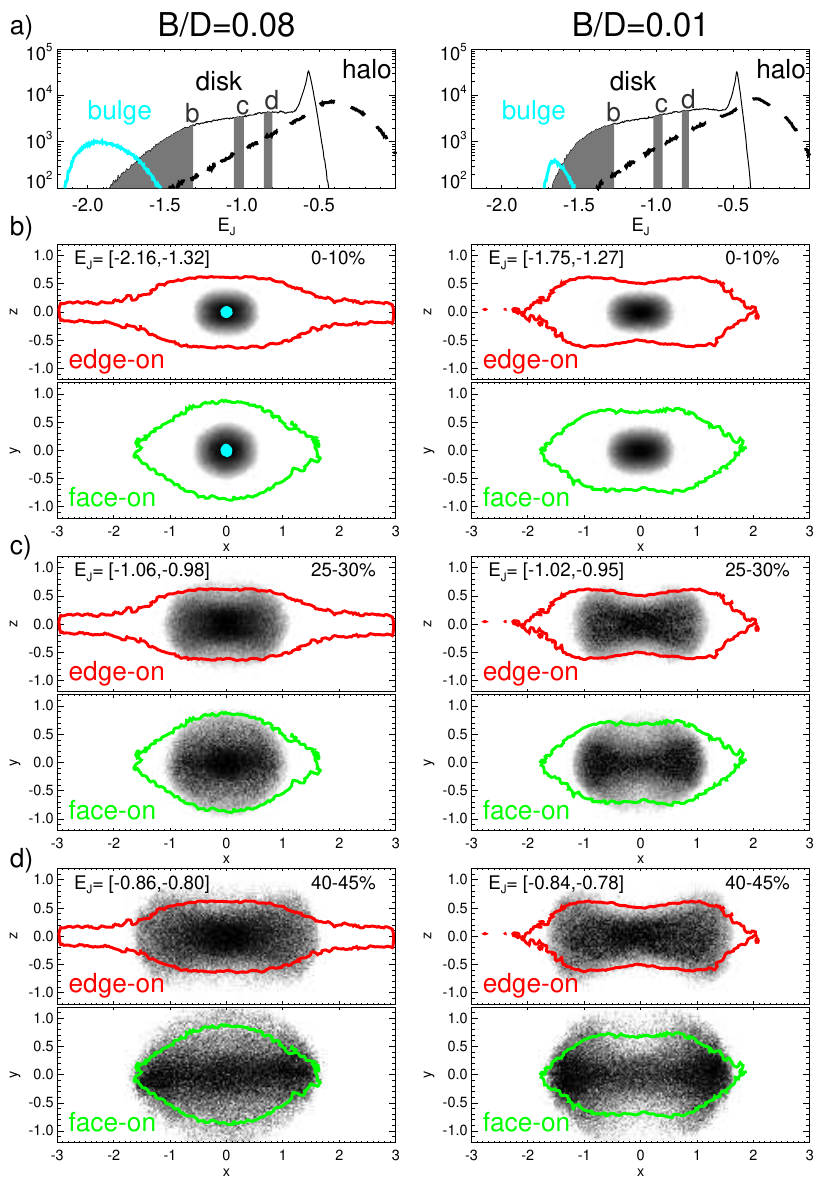}
\caption{Orbital structure in $B/D=0.08$ {\em (left column)} and
  $B/D=0.01$ {\em (right column)} simulations. The uppermost row a)
  shows the distribution of Jacobi energy $E_J$: the bulge, disk, and
  halo components are shown separately. The frames b)-d) show the 
    time-averaged (between 10-11 Gyrs) density of orbits, for the disk
    particles falling in the three $E_j$ ranges indicated by the
    shaded areas in the histograms of a): the upper frame gives an
    edge-on view and the lower frame a face-on view.  Thick contours
    outline the edge-on (red) and face-on (green) bar morphology. The
    blue filled circle in b) indicates where the projected surface
    density of the classical bulge exceeds that of the disk.
\label{fig_simu_orbits}
}
\end{figure}

The difference in the orbital structure of the two above
  simulations is illustrated in Fig. 3, showing the time-averaged
  density distributions of selected disk particle orbits, in a frame
  co-rotating with the bar.  The plots were constructed by sampling
  the positions of all particles at 500 equally separated instants
  during the time interval $T=10-11$ Gyrs, when the bar pattern speed
  $\Omega_{bar}$ had attained a constant value (within a few
  percents). The bar pattern speed was measured by calculating the
  moment of inertia tensor for all disk particles within 3 distance units from
  the center, and following the direction of its longest principal
  axis: the mean rate of change over the time interval was used as
  $\Omega_{bar}$. The positions of particles were converted to the
  rotating system and tabulated in two-dimensional bins in xy and
  xz-planes, where x denotes the coordinate along the bar major
  axis. The tabulations from all 500 sampling times were added
  together, yielding the face-on and edge-on projections of the
  time-averaged density of the selected orbits.  We also calculated
  the Jacobi energies of the particles, $E_j= \frac{1}{2} v^2 + \Phi
  -\frac{1}{2} {\Omega_{bar}}^2 (x^2+y^2)$, where $\Phi$ is the
  gravitational potential and $v$ the velocity in the rotating frame:
  over the sampling interval the $E_j$ of particles stayed nearly
  constant.

In Fig. 3, the orbits of disk particles in three different ranges of
Jacobi energy are shown separately in frames b), c), and d); the ranges
are indicated at the $E_j$ histogram in frame a).  In the $B/D=0.08$
run (left frames), the orbits of the disk particles with $E_J$ in the
lowest 10\% range (frame b, left) fill a nearly circular region in the
face-on projection. For the same $E_J$ range, in the $B/D=0.01$ run
(frame b, right) the density distribution is clearly elongated along
the bar major axis.  Similar difference concerns the orbits in the
energy range $25-30\%$ (frame c), which more or less outlines the
circular barlens in the $B/D=0.08$ case, and in the energy range
$40-45\%$ (frame d), comprising of the orbits with the largest
vertical extent. For the $B/D=0.01$ model the orbits in frames c) and d)
show a clear box-orbit character, with pronounced narrowing in the bar
minor axis direction.  Clearly, a significant central mass
concentration is able to affect the orbital structure in a significant
portion of the bar, favoring loop like orbits responsible for the
round face-on morphology.

The fundamental difference in the above $B/D=0.08$ and $B/D=0.01$
simulations is the much higher central density in the former model,
leading to significantly steeper inner rotation curve. The effect of
the inner rotation curve is further illustrated in Fig. 4 
  (uppermost frame), displaying a range of $B/D$ ratios from $0.005$
  to $0.16$, while the bulge effective radius is kept fixed to $r_{\rm
    eff}=0.07 h_r$.  In these experiments a hint of X-shape face-on
  morphology (manifests in isophotes as a waist in the minor axis; see
  the insert figures) appears when the central slope $dV_{cir}/dr
  \lesssim 5 V_{max}/h_r$, taking place for $B/D \lesssim 0.02$.  The
  same threshold applies to models where the inner slope is reduced by
  increasing the bulge effective radius: for example, doubling the
  effective radius to $r_{\rm eff}=0.13 h_r$ for $B/D=0.08$, yields a
  clear X-signature (middle frame).  Similar dependence on rotation
  curve slope is seen also in models which have no classical bulge
  component, when the degree of halo central concentration is varied
  (lowermost frame).  Note that the rotation curve is displayed at the
  time when the bar has formed ($T \sim 5$ Gyrs); the initial $h_r$ is
  used here mainly to get a rough normalization of the slopes: during
  the formation of the bar the outer disk scale length typically
  increases by a factor of $50\%-100\%$.  (see Fig. 1).

Since the simulated morphology is very sensitive to the steepness
  of the central velocity slope, even the use of too large gravity
  softening may prevent the pure barlens morphology. This is
  illustrated in Fig. 5, for the concentrated halo model ('halo3') of
  Fig. 4, which with our nominal softening ($\epsilon/h_r =0.01$)
  leads to barlens morphology. Doubling the softening value
  ($\epsilon/h_r = 0.02$) leads to clearly elongated central
  structure. In contrast, reducing $\epsilon/h_r$ to 0.005 yields
  practically similar  pure barlens morphology as the nominal
  value. For our model with $B/D=0.08$, a similar change of morphology
  takes place between $\epsilon/h_r = 0.02$ and $0.05$.
  Interestingly, the edge-on morphology is much less affected by
  increased softening: since the attention in most previous
  simulations of B/P/X structure has concentrated on the bar vertical
  structure, the dependence of face-on morphology on softening may
  have escaped attention.

\begin{figure}[ht!]
\figurenum{4}
\epsscale{1.2}
\plotone{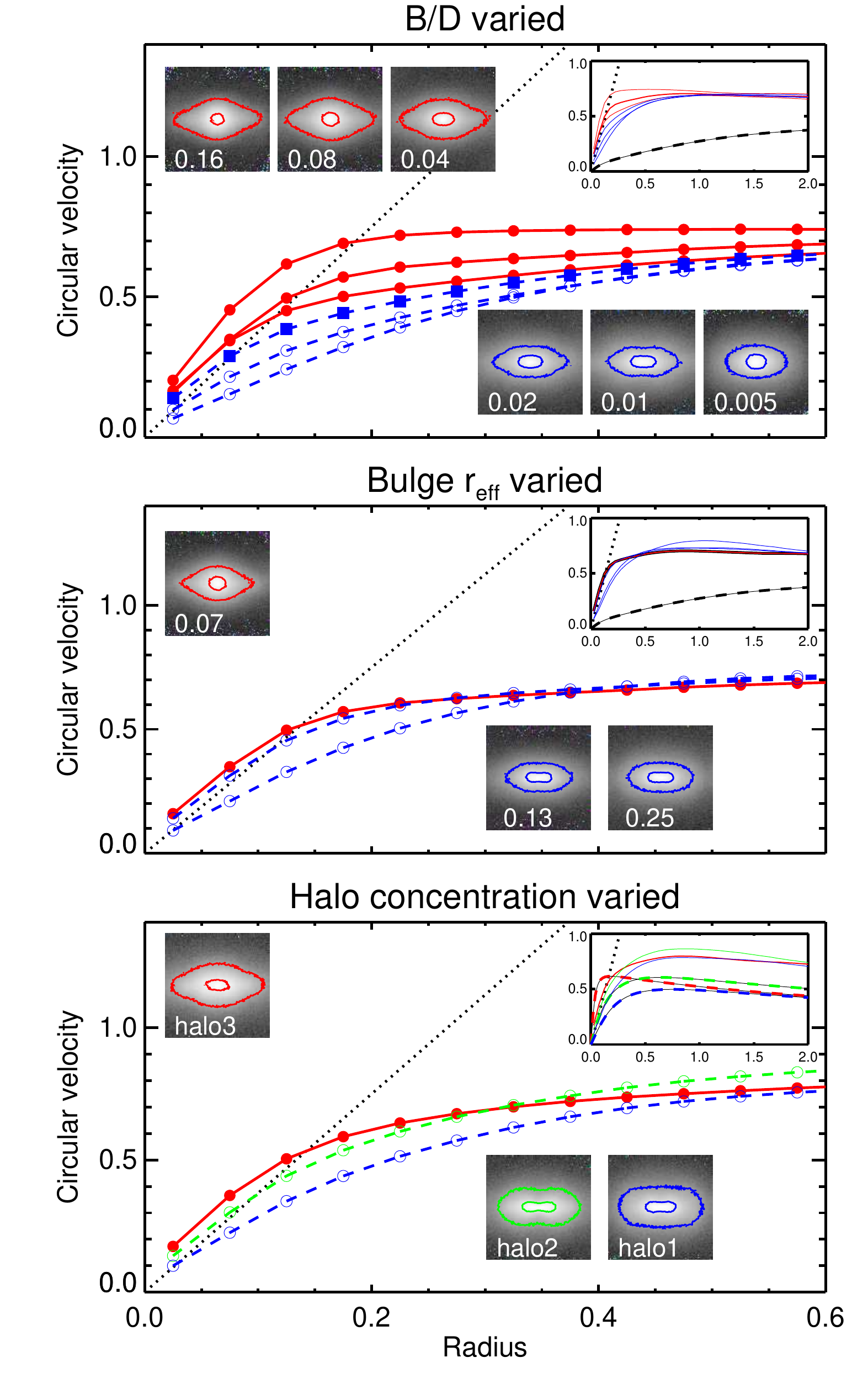}
\caption{The central slope of circular velocity curve in
  simulations with different $B/D=0.005-0.16$ ({\em uppermost frame};
  $r_{\rm eff}/h_r$ fixed to 0.07) and different bulge $r_{\rm
    eff}/h_r =0.07-0.25$ ({\em middle frame}; $B/D$ fixed to 0.08),
  and in bulgeless simulations with different degree of halo
  concentration ({\em lowermost frame}; halo concentration increases
  from halo model 'halo1' to 'halo3'); red curves indicate simulations
  leading to a pure barlens-morphology (round inner morphology with no
  trace X on face-on view). The dashed line indicates the approximate
  slope separating the morphologies ($dV_{cir}/dr = 5
  V_{max}/h_r$). The insert figures display the rotation curve on a
  larger scale, with halo contribution marked with a dashed line; the
  insert snapshots show the barlens morphology (here using only the
  disk particles).  The rotation curve corresponds to T=5 Gyrs;
  however a practically similar threshold would apply when using the
  rotation curves from the time before the bar has formed.
  \label{fig_simu_slopes}
}
\end{figure}

\begin{figure}[ht!]
\figurenum{5}
\epsscale{1.2}
\plotone{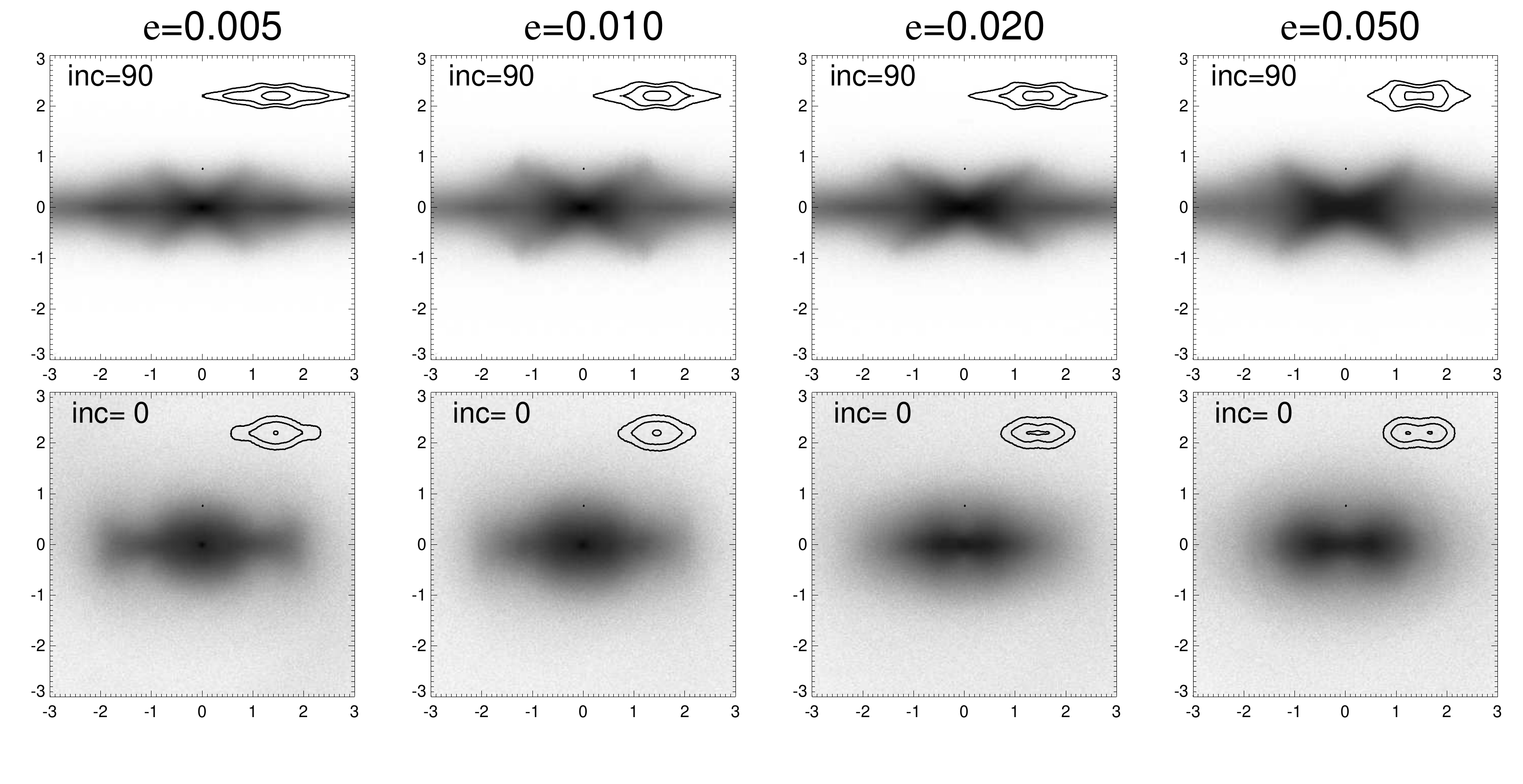}
\caption{ The effect of softening in the simulated inner morphology
  (total extent of the frames is 6 simulation units). From left to right the
  softening increases from $\epsilon/h_r=0.005$ to $0.05$. Even a
  2-fold softening compared to the nominal value $0.01$ destroys the
  barlens morphology, while with a smaller softening an even slightly
  better resemblance to observed pure barlens morphology is attained.
  \label{fig_simu_eps}
}
\end{figure}

\begin{figure}[ht]
\figurenum{6}
\epsscale{1.2}
\plotone{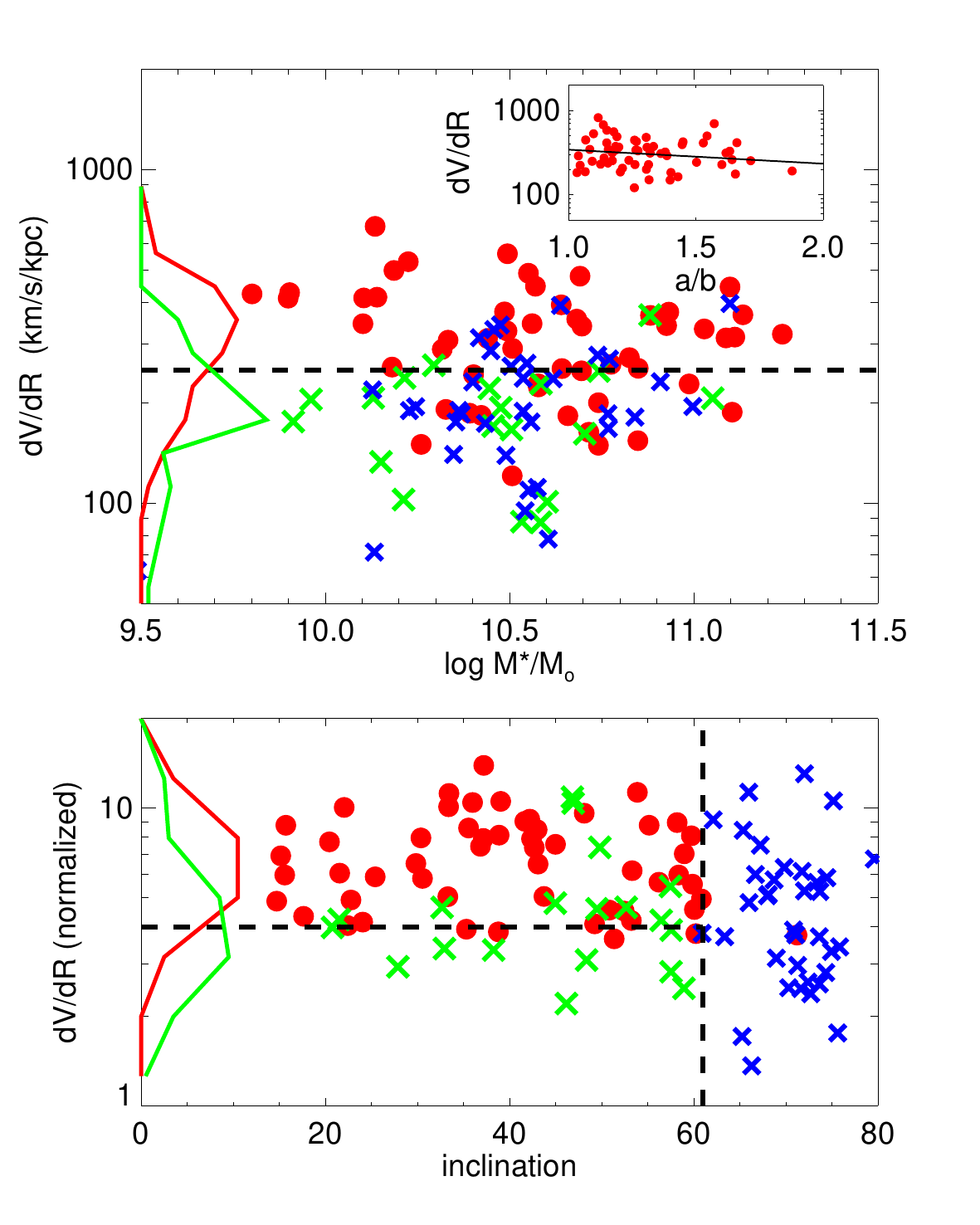}
\caption{Slopes of the inner rotation curves for S$^4$G galaxies,
  calculated from the 3.6$\mu$m images.  The red circles indicate
  galaxies with a barlens identification \citep{laurikainen2011,
    buta2015}, while green ($i<60^\circ$) and blue crosses
  ($i>60^\circ$) stand for galaxies with an X-signature. In the {\em
    upper frame} the inner slope is shown versus galaxy stellar mass
  ($M^*$ from \cite{munozmateos2015}), while in the {\em lower frame}
  the slope is normalized by $V_{max}/h_r$ and displayed versus galaxy
  inclination. Here $V_{max}$ is the calculated maximum velocity
    due to visible mass and $h_r$ is the outer disk scale length from
    \cite{salo2015}.  All parameter values used in Fig. 6 are also
    listed in Table 1.  Histograms in the left show the distributions
  of barlens and X-shape parent galaxies: they cover all
    inclinations, so that both prominent edge-on X shapes and weaker
    X-like signatures at small and intermediate inclinations are
    included. The insert in the upper frame shows the central
  velocity slope vs barlens axial ratio (deprojected to the disk
  plane).
\label{fig_s4g_slopes}
}
\end{figure}

\section{Comparison to observations} \label{sec:obs}

Above simulations suggest that there should be a more or less
clear-cut difference (without/with trace of X) in the inner morphology for
barlens galaxies with steep/shallow inner rotation curve slopes,
respectively. Can this be verified with observations?  Unfortunately,
the number of barlens galaxies with reliably measured rotation curves
is very small: for example, the recent study of inner rotation curves
based on high resolution $H_\alpha$ kinematics of 29 S$^4$G galaxies
\citep{erroz2016} contains only two galaxies in common with the
compilation of barlens \& X-shape galaxies in \cite{laurikainen2014}.
  However, \citet{erroz2016} also conclude that the inner
rotation curves are dominated by  baryonic matter (see also \cite{lelli2013}), based on the
comparison of inner slopes of $H_\alpha$ rotation curves with those derived
from the $S^4$G mass maps \citep[non-stellar contaminants removed based
  on both 3.6 $\mu$m and 4.5 $\mu$m images;][]{querejeta2015} using
the above mentioned NIRQB code.  We may therefore approximate the
inner rotation curve slopes with those derived from the mid-IR images: this
is done in Fig. 6 for all barlens \& X-shape galaxies in S$^4$G.
Since the rotation curves calculated from the raw 3.6 $\mu$m S$^4$G
images and the mass maps differ only very little \citep{diaz2016a},
the slopes in the figure are based on raw 3.6 $\mu$m images.  The
median ratio between the observed slopes in \citet{erroz2016} and our
estimated slopes is 0.96 (using the 24 galaxies with observed slopes given in Table 3
of \citet{erroz2016}).

Figure 6 (upper frame) displays the inner rotation curve slopes (in
units of km/sec/kpc) versus galaxy stellar mass. The slopes of barlens
galaxies are on average a factor of two steeper than those with
X-shape structures.  Also, the insert figure indicates that there
might be a tendency for a barlens to be rounder when the central slope
is steeper, which is in accordance with the simulated behavior.
Although the fit is not statistically significant, this dependence on
slope is interesting since in \cite{laurikainen2014} and
\cite{athanassoula2015} no trend was found between the axial ratio of
the barlens and the bar strength.  Since the visibility of possible
weak X-signatures depends strongly on observing direction
\citep{laurikainen2016b}, the lower frame makes a comparison as a
function of galaxy inclination. Also, to eliminate the influence of
galaxy mass, the slopes are normalized by $V_{max}/h_r$.
Excluding the highly inclined galaxies with $i \gtrsim 60^\circ$ 
  (which all are X-shaped), the separation between barlens and
X-shape parent galaxies is strikingly clear, in good qualitative
agreement with the simulation predictions. In the figure the
  division between pure barlens and low-inclination X-shape is drawn
  at 250 km/s/kpc and at 4 $V_{max}/h_r$. The latter is close to
  simulation prediction; note however that here $V_{max}$ refers to
  stellar-only contribution, and $h_r$ is the observed disc scale
  length - after the bar has formed.

 Three galaxies with X-shaped bars, NGC 4123, NGC 4725 and NGC
  7140, appear in the barlens region in Fig. 6, lower frame, where the
  normalized $dV_{cir}/dr$ is used (they have $i<60^\circ$ and
  $dV_{cir}/dr>8 V_{max}/h_r$; see Table 1). For NGC 4725 this is the case also
  when expressing the central slope in km/s/kpc (upper frame). In NGC
  4725 most probably a large fraction of the central mass
  concentration inside the X-feature comes from a nuclear bar.  In NGC
  7140 it is largely due to a nuclear ringlens, and in NGC 4123 due to
  a point-like nucleus. All these nuclear features are expected to
  have formed secularly from the disk material. There is no clear
  indication why these three galaxies deviate from the common trend; this
  might be related to factors not addressed by the simulations, like
  the gas content or the halo-to-disk mass ratio.

The morphology of the above simulation models is further studied in
\cite{laurikainen2016b}, where we demonstrate that the $B/D=0.08$
model, when viewed from different inclinations and orientations with
respect to the bar major axis, traces well the observed distribution
of apparent sizes (normalized to the bar size) and aspect ratios of
barlens and X features versus parent galaxy inclination.  The same
model also reproduces the observed dependence of barlens boxiness
parameter on galaxy inclination: for $i \lesssim 40^\circ$ barlenses
have $B_4 > 0$, but for larger inclination $B_4$ gets negative,
i.e. isophotes become boxy. This is in line with \cite{erwin2013} who suggested
that B/P bulges can be identified at intermediate inclinations
($i=40-60^\circ$) by the boxy inner isophotes of bars.

\section{Discussion} \label{sec:discussion}

The above simulations and comparisons to observations suggest that a
steep inner rotation curve slope favors a pure barlens
face-on morphology, characterized by a nearly round central appearance.
With a shallower rotation curve the orbits in the central regions become
more elongated along the bar major axis, leading to an oval shaped
bar, or even to an X-shape face-on feature (narrow waist in the minor axis). 

The strong constrain on the simulated inner slope is likely to explain
why barlenses, which are fairly conspicuous observationally, have not
been encountered in simulation studies before \citet{laurikainen2014}
and \citet{athanassoula2015}: typically the classical bulge components
employed, if included at all, have been less centrally concentrated
than in the current simulations.  In \cite{debattista2004} the problem
of not attaining round bar-related bulges in simulations was
acknowledged, and it was speculated that extended central objects with
$B/D = 0.1-0.2$ might be needed, based on \cite{shen2004} simulations
studying the robustness of bars against central black hole
masses. In \cite{shen2004} simulations a rigid halo was assumed,
  however, very similar conclusions were reached in
  \cite{athanassoula2005} using live halos.  Our current simulations
demonstrate that a substantially smaller $B/D$ (of a few percents) are
sufficient, provided that realistic small effective bulge radii from
observations ($r_{\rm eff}/h_r \lesssim 10\%$) are adopted. Such
central mass concentrations are able to affect the orbital morphology
up to $\sim 10 r_{\rm eff}$, while still not endangering the survival
of the narrow bar.  Simulating steep central slopes of course implies
a need for small gravity softening: in particular, cosmological
simulations so far lack the resolution necessary to resolve the inner
bar dynamics \citep[see e.g. the discussion in ][]{brooks2016}.

\begin{figure}[ht!]
\figurenum{7}
\epsscale{1.2}
\plotone{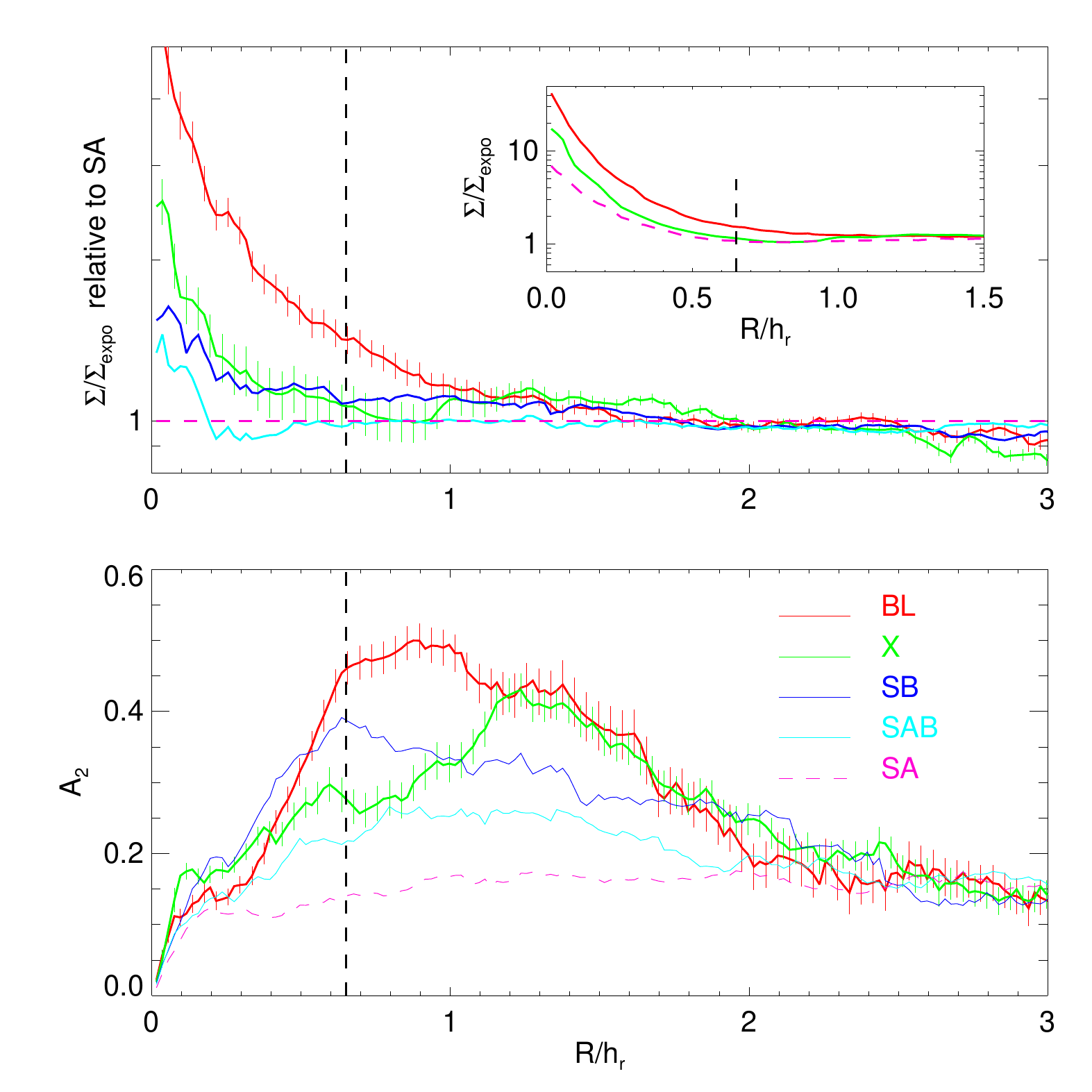}
\caption{The insert in the {\em upper frame} shows the median-stacked
  surface brightness excess of barlens, X-shape, and SA galaxies,
  relative to their exponential outer disk profiles extrapolated to
  the center. Decomposition parameters for the disks are from
  \citet{salo2015} and the bar family classification from
  \cite{buta2015}.  In the large frame the excess brightness of
  barlens (bl) and X samples is shown relative to SA galaxies. For
  comparison also samples of SB and SAB are included (all samples are
  limited to galaxies with $M^*>10^{10}M_\sun, T \le 5$). The {\em
    lower frame} indicate the medians of $A_2$ Fourier-amplitude
  profiles. The vertical dashed line indicates the median barlens size relative to disk scale length \cite[$r/h_r \approx 0.65$; from][]{herrera2015}.
\label{fig_ratio_a2}}
\end{figure}

We have demonstrated that an inclusion of a small classical bulge in
simulations can reproduce the barlens morphology. Does this imply that
the observed barlens galaxies harbor classical bulges which were in
place already before the bar formed?  An alternative is that the
central mass concentrations inside the barlens region have
formed together with the bar, via bar-induced inflow and star
formation.  Namely, barlenses are typically associated with strong
bars, and to galaxies which have consumed most of their gas.  Such
inflow was also the mechanism that created the central mass
concentrations in the \cite{athanassoula2015} simulations, 
which did not include any initial classical bulge component.
Although most of the simulated galaxies had traces of X-shape in their face-on morphology, it is
feasible that a somewhat stronger inflow  in their models could
  have created even steeper central rotation slopes and thus more
  realistic pure barlens morphologies.  

Observational evidence for the possible bar-related growth is provided
by Fig. 7, which compares the amount of central density excess in
barlens galaxies to other types of barred galaxies (with X, and with
those SB, SAB galaxies which do not have barlens or X morphology), and
to non-barred SA galaxies. Clearly, barlens galaxies have the
strongest central excess (consistent with their early $T$ types), both
at the region corresponding to the barlens itself, $r/h_r \sim
0.5-0.8$, and in the very center, $r/h_r <0.2$.  Among the different
galaxy families, the central excess correlates with the strength of
the bar (or non-axisymmetric perturbation in general, here measured in
terms of $A_2$ Fourier amplitudes of surface density), supporting the
role of bars in building the central concentration. Similar conclusion
was recently reached in \cite{diaz2016b} when comparing SB, SAB, and
SA galaxies: however the difference is far more pronounced when
barlens galaxies are considered.  

{Nevertheless, the above link between strong bars and central
  concentrations still leaves open the possibility that the observed
  central components represent classical bulges which have helped the
  bars to gain strength, by removing the angular momentum of the bar.
  Such dependence was seen e.g. in the simulations of
  \citep{atha2002}, where the bar $A_2$ maximum was nearly doubled in
  a model that included a classical bulge component, compared to a
  model with only disk and halo components. However, the classical
  bulge in those simulation was quite extended and massive, with $B/D
  = 0.60$. In the current simulations with small compact classical
  bulges the strength of the bar is practically independent of the
  adopted initial $B/D$, so that no similar straight-forward
  conneection can be made.  Clearly, spectroscopic observations of the
  stellar ages and metallicities would be needed to distinguish
  between the disky pseudo-bulge/classical bulge character of the
  central concentrations in barlens galaxies.}

\section{Conclusions}

The barlens-sized central light concentrations in early-type
disk galaxies are often identified as massive classical bulges. In
this paper we have provided further evidence that, at least in the
case of barred galaxies, they represent B/P/X/bl structures related to
secular evolution of bars.  However, the exact morphology at low
galaxy inclination depends on the galaxy mass distribution.

Our main conclusions are the following:

 1) Stellar dynamical simulations, using parameter values appropriate
 for MW mass galaxies, indicate that a steeply rising inner rotation
 curve is required for obtaining a pure barlens morphology (no trace
 of X in low inclination view). In the current simulations this was
 achieved by including a small classical bulge to the initial values,
 or by using a sufficiently centrally concentrated halo.  The
 threshold inner slope in simulations is $dV_{cir}/dr \sim 5 V_{max}/h_r$.

2) The simulated behavior is in qualitative agreement with slopes
derived from $S^4$G 3.6 $\mu$m images, which indicate that barlens
galaxies have steeper baryonic rotation curves ($dV_{cir}/dr \gtrsim 250$
km/sec/kpc) than the galaxies which exhibit X-signature even at
$i<60^\circ$.


3) Comparisons of stacked surface density profiles indicate that
barlens galaxies have larger $A_2$ amplitudes than any other type of
barred galaxies.  They also have the largest relative excess of inner
surface density (within $r/h_r <  0.2$), which can be interpreted as evidence for bar induced inflow in
the past.

\section{Acknowledgments}

We thank the referee for useful suggestions on how to clarify the
presentation of the results.
This work was supported by the {\em DAGAL} network: People Programme
(Marie Curie Actions) of the European Unions Seventh Framework
Programme FP7/2007-2013 under REA grant agreement number
PITN-GA-2011-289313. The grant from Academy of Finland (n:o 297738) is
also acknowledged.

\onecolumngrid

\newpage

\begin{deluxetable}{llcccccc}
\tablecaption{Data for  galaxies with barlens and X features used in Fig. 6.}
 \tablehead{Galaxy & Hubble Type & $dV_{cir}/dr$   & $dV_{cir}/dr$ & $V_{max}$  &  $h_r$ & INC     & $\log_{10}(M^*/M_\sun)$ \\ 
                   &             & km/s/kpc & normalized &km/s      &  kpc   & degrees & } \startdata
\sidehead{ Galaxies with barlenses}\\
IC2051     & SB($\underline{\rm r}$s,bl)b                                                     &    223 &   4.20 &    155 &   2.92 &   53.2 &  10.58 \\
NGC0613    & SB($\underline{\rm r}$s,bl,nr)b                                                  &    312 &   8.12 &    211 &   5.50 &   38.9 &  11.09 \\
NGC0936    & (L)SB$_{\rm a}$($\underline{\rm r}$s,bl)0$^+$                                    &    340 &   7.90 &    188 &   4.37 &   42.4 &  10.93 \\
NGC1015    & (R$^{\prime}$)SB(r,bl)0/a                                                        &    186 &   5.81 &    123 &   3.87 &   30.5 &  10.39 \\
NGC1022    & (RL)SAB($\underline{\rm r}$s,bl,ns)$\underline{\rm 0}$/a                         &    498 &   6.05 &    153 &   1.86 &   21.6 &  10.19 \\
NGC1079    & (R$\underline{\rm L}$)S($\underline{\rm A}$B$_{\rm a}$($\underline{\rm r}$s,bl)0 &    227 &   4.52 &    145 &   2.88 &   52.4 &  10.58 \\
NGC1097    & (R$^{\prime}$)SB(rs,bl,nr)ab pec                                                 &    321 &   9.63 &    265 &   7.96 &   48.1 &  11.24 \\
NGC1300    & (R$^{\prime}$)SB(s,bl,nrl)b                                                      &    227 &  10.12 &    122 &   5.45 &   33.4 &  10.58 \\
NGC1302    & (RLRL)SAB($\underline{\rm r}$l,bl)0$^+$                                          &    345 &   8.77 &    153 &   3.89 &   15.7 &  10.56 \\
NGC1326    & (R$_1$)SAB$_{\rm a}$(r,bl,nr)0$^+$                                               &    488 &   7.92 &    190 &   3.09 &   37.2 &  10.55 \\
NGC1350    & (R)SAB$_{\rm a}$(r,bl)0/a                                                        &    273 &   8.95 &    158 &   5.18 &   58.2 &  10.82 \\
NGC1398    & (R$^{\prime}$R)SB($\underline{\rm r}$s,bl)a                                      &    367 &   9.19 &    222 &   5.57 &   42.2 &  11.13 \\
NGC1452    & (RL)SB($\underline{\rm r}$s,bl)0/a                                               &    242 &   6.17 &    133 &   3.38 &   53.3 &  10.40 \\
NGC1512    & (R$\underline{\rm L}$)SB(r,bl,nr)a                                               &    308 &   7.37 &    116 &   2.79 &   42.7 &  10.33 \\
NGC1533    & (RL)SB(bl)0$^{\circ}$                                                            &    447 &   5.97 &    175 &   2.34 &   15.6 &  10.57 \\
NGC1640    & (R$^{\prime}$)SB$_{\rm a}$(r,bl)ab                                               &    255 &   4.14 &    114 &   1.85 &   24.1 &  10.18 \\
NGC2543    & SAB(s,bl)b                                                                       &    183 &   5.56 &    119 &   3.61 &   59.9 &  10.42 \\
NGC2787    & (L)SB$_{\rm a}$(r,bl)0$^{\circ}$                                                 &    529 &   5.64 &    148 &   1.59 &   56.2 &  10.23 \\
NGC2859    & (R)SAB$_{\rm a}$(rl,bl,nl,nb)0$^+$                                               &    365 &  13.93 &    215 &   8.20 &   37.2 &  10.88 \\
NGC2968    & (L)SB(s,bl)0$^+$                                                                 &    412 &   6.49 &    126 &   1.99 &   43.1 &  10.10 \\
NGC3351    & (R$^{\prime}$)SB(r,bl,nr)a                                                       &    374 &   7.57 &    140 &   2.85 &   45.0 &  10.49 \\
NGC3384    & (L)SA$\underline{\rm B}$(bl)0$^-$                                                &    328 &   4.96 &    156 &   2.37 &   60.8 &  10.49 \\
NGC3489    & (R)SA$\underline{\rm B}$(r,bl)0$^{\circ}$:                                       &    676 &   4.57 &    146 &   0.99 &   60.1 &  10.14 \\
NGC3941    & (R)SB$_{\rm a}$(bl)0$^{\circ}$                                                   &    559 &   4.55 &    179 &   1.46 &   50.8 &  10.49 \\
NGC3953    & SB(r,bl)b                                                                        &    227 &   5.97 &    162 &   4.25 &   58.4 &  10.99 \\
NGC3992    & SB(rs,bl,nb)ab                                                                   &    187 &   8.77 &    173 &   8.10 &   55.1 &  11.10 \\
NGC4245    & (RL)SB(r,bl,n$\underline{\rm r}$l)0$^+$                                          &    423 &   5.05 &     95 &   1.13 &   33.3 &   9.80 \\
NGC4314    & (R$_1$')SB(r$\underline{\rm l}$,bl,nr)a                                          &    414 &   7.71 &    118 &   2.20 &   20.4 &  10.14 \\
NGC4371    & (L)SB$_{\rm a}$(r,bl,nr)0$^+$                                                    &    291 &   7.03 &    145 &   3.52 &   59.0 &  10.51 \\
NGC4394    & ($\underline{\rm R}$L)SB(rs,bl,nl)0/a                                            &    311 &   7.95 &    116 &   2.97 &   30.4 &  10.44 \\
NGC4448    & (R)SB(r,bl)0/$\underline{\rm a}$                                                 &    153 &   3.75 &    162 &   3.95 &   71.2 &  10.85 \\
NGC4454    & (RL)SAB(r,bl)0/a                                                                 &    182 &   4.33 &    143 &   3.39 &   17.6 &  10.66 \\
NGC4548    & SB(rs,bl)$\underline{\rm a}$b                                                    &    339 &  10.55 &    141 &   4.39 &   39.0 &  10.70 \\
NGC4579    & ($\underline{\rm R}$LR')SB(rs,bl)a                                               &    444 &   9.04 &    211 &   4.30 &   41.6 &  11.10 \\
NGC4593    & (R$^{\prime}$)SB(rs,bl,AGN)a                                                     &    373 &  11.22 &    212 &   6.38 &   33.4 &  10.93 \\
NGC4596    & (L)SB(rs,bl)$\underline{\rm 0}$/a                                                &    356 &   8.59 &    159 &   3.84 &   35.5 &  10.68 \\
NGC4639    & (R$^{\prime}$)SA$\underline{\rm B}$(rs,bl)ab                                     &    289 &   4.09 &    134 &   1.90 &   49.2 &  10.32 \\
NGC4643    & (L)SB($\underline{\rm r}$s,bl,nl)0$^+$                                           &    332 &   7.45 &    228 &   5.11 &   36.8 &  11.03 \\
NGC4754    & (L)SB$_{\rm a}$(bl)0$^{\circ}$                                                   &    393 &   8.08 &    167 &   3.43 &   59.7 &  10.64 \\
NGC4795    & (R$^{\prime}$)SA$\underline{\rm B}_{\rm a}$(l,bl)a pec                           &    200 &   5.05 &    157 &   3.98 &   43.7 &  10.74 \\
NGC4984    & (R$^{\prime}$R)SAB$_{\rm a}$(l,bl,nl)0/a                                         &    479 &  11.31 &    217 &   5.14 &   53.9 &  10.69 \\
NGC5101    & (R$_1$R$_2$')SB($\underline{\rm r}$s,bl)0/a                                      &    314 &  10.07 &    218 &   6.99 &   22.0 &  11.11 \\
NGC5134    & (R)SAB(rs,bl)a                                                                   &    411 &   4.87 &    103 &   1.23 &   14.7 &   9.90 \\
NGC5339    & SA$\underline{\rm B}$(rs,bl)ab                                                   &    120 &   3.84 &    123 &   3.94 &   38.8 &  10.51 \\
NGC5347    & SB(rs,bl)a                                                                       &    427 &   4.92 &    174 &   2.01 &   22.8 &   9.90 \\
NGC5375    & (R$^{\prime}$)SB$_{\rm a}$(rs,bl)$\underline{\rm a}$b                            &    163 &   6.52 &    138 &   5.54 &   29.8 &  10.72 \\
NGC5701    & (R$_1$')SA$\underline{\rm B}$(rl,bl)0/a                                          &    249 &   6.92 &    162 &   4.51 &   15.2 &  10.69 \\
NGC5728    & (R$_1$)SB($\underline{\rm r}^{\prime}$l,bl,nr,nb)0/a                             &    253 &   8.47 &    179 &   6.01 &   43.0 &  10.85 \\
NGC5750    & (RL)SAB(r$^{\prime}$l$\underline{\rm r}$,s,bl)0/a                                &    149 &   3.79 &    153 &   3.90 &   60.2 &  10.74 \\
NGC5770    & SAB($\underline{\rm r}$l,bl)0$^+$                                                &    345 &   4.03 &    132 &   1.55 &   22.4 &  10.10 \\
NGC5850    & (R$^{\prime}$)SB(r,bl,nr,nb)$\underline{\rm a}$b                                 &    260 &  10.46 &    139 &   5.59 &   36.0 &  10.77 \\
NGC5957    & (R$^{\prime}$)SA$\underline{\rm B}$(rs,bl)$\underline{\rm a}$b                   &    150 &   5.88 &    111 &   4.37 &   25.4 &  10.26 \\
NGC6014    & SAB(rs,bl)$\underline{\rm 0}$/a                                                  &    191 &   3.92 &    119 &   2.45 &   35.3 &  10.33 \\
NGC7079    & (L)SA$\underline{\rm B}_{\rm a}$(s,bl)0$^{\circ}$:                               &    253 &   3.64 &    165 &   2.38 &   51.3 &  10.64 \\
\sidehead{Galaxies with X}\\
ESO404-027 & SAB(s)$\underline{\rm a}$b:                                                      &     71 &   3.14 &     78 &   3.43 &   69.0 &  10.13 \\
IC1067     & SB(r,bl)b                                                                        &    175 &   3.34 &     91 &   1.74 &   38.3 &   9.91 \\
IC3806     & SA(r)0$^+$                                                                       &     62 &   1.71 &     65 &   1.78 &   65.2 &   9.49 \\
IC4237     & SB(r)b                                                                           &     87 &   2.21 &    135 &   3.41 &   46.2 &  10.53 \\
IC5240     & SB$_{\rm x}$(r)0/a                                                               &    170 &   4.80 &    122 &   3.44 &   45.0 &  10.45 \\
NGC0532    & SAB$_{\rm xa}$(r)0/a                                                             &    109 &   3.70 &    118 &   4.01 &   73.6 &  10.55 \\
NGC0615    & (R')SA$_{\rm x}$(r)$\underline{\rm a}$b                                          &    236 &   4.82 &    149 &   3.05 &   66.0 &  10.54 \\
NGC0779    & (L)SA$_{\rm x}$(rs)a                                                             &    175 &   2.50 &    130 &   1.86 &   70.3 &  10.35 \\
NGC1461    & S$\underline{\rm A}$B(r)0$^{\circ}$                                              &    231 &   3.90 &    131 &   2.21 &   70.8 &  10.40 \\
NGC1476    & Im sp                                                                            &     40 &   1.36 &     44 &   1.48 &   66.3 &   9.14 \\
NGC2654    & SB$_{\rm x}$(r,nd)0/a sp                                                         &    175 &   2.81 &    146 &   2.35 &   74.3 &  10.56 \\
NGC2683    & (R'L)SB$_{xa}$(rs)0/a sp                                                         &    236 &   3.30 &    163 &   2.29 &   74.9 &  10.62 \\
NGC3185    & (RL)SAB$_{\rm ax}$(rs,bl)a                                                       &    237 &   4.59 &    115 &   2.23 &   49.5 &  10.22 \\
NGC3254    & SA$\underline{\rm B}_{xa}$b                                                      &    111 &   5.64 &    106 &   5.40 &   73.3 &  10.57 \\
NGC3301    & (R'L)SAB$_{\rm x}$(r)0$^+$ sp                                                    &    286 &   6.13 &    139 &   2.99 &   71.7 &  10.45 \\
NGC3380    & (R$\underline{\rm L}$)SAB($\underline{\rm r}$s,bl)0/a                            &    204 &   3.98 &    102 &   1.99 &   20.8 &   9.96 \\
NGC3424    & S$_{\rm x}\underline{\rm a}$b: sp pec                                            &    173 &   3.42 &    143 &   2.83 &   75.8 &  10.43 \\
NGC3623    & (R')SAB$_{\rm x}$(rs)a                                                           &    231 &   5.25 &    175 &   3.98 &   73.7 &  10.91 \\
NGC3673    & (R')SA$\underline{\rm B}_{\rm x}$(rs)ab                                          &    166 &   4.65 &    122 &   3.41 &   52.6 &  10.51 \\
NGC3692    & (R'L)SA(r)0/a sp                                                                 &     94 &   1.76 &    135 &   2.51 &   75.6 &  10.54 \\
NGC3887    & (RL)SA$\underline{\rm B}_{\rm x}$(rs)bc                                          &    193 &   3.38 &    145 &   2.54 &   32.9 &  10.48 \\
NGC4123    & SB$_{\rm x}$(rs)a$\underline{\rm b}$                                             &    259 &  10.42$^*$ &     99 &   3.99 &   46.9 &  10.29 \\
NGC4192    & (R$_{1}$')SAB$_{\rm x}$(rs,nd)$\underline{\rm a}$b                               &    270 &  13.06 &    140 &   6.79 &   72.0 &  10.77 \\
NGC4216    & (R$_{2}$')SAB$_{\rm ax}$(r,nd)$\underline{\rm a}$b sp/E7-8                       &    194 &   6.76 &    158 &   5.50 &   79.6 &  11.00 \\
NGC4220    & (L)SAB(r)0$^+$                                                                   &    189 &   2.61 &    130 &   1.79 &   72.4 &  10.36 \\
NGC4235    & S$_{\rm x}$0$^+$ sp                                                              &    257 &   5.28 &    135 &   2.77 &   72.0 &  10.50 \\
NGC4268    & S$\underline{\rm A}$B($\underline{\rm r}$s)0$^+$: sp                             &    194 &   3.80 &    121 &   2.38 &   60.9 &  10.24 \\
NGC4293    & $\underline{\rm R}$(L)SB$_{\rm x}$(r$\underline{\rm s}$)0/a                      &    313 &   9.15 &    119 &   3.48 &   62.1 &  10.42 \\
NGC4419    & SA$\underline{\rm B}_x$0/$\underline{\rm a}$ sp/E6                               &    331 &   2.96 &    165 &   1.48 &   71.3 &  10.46 \\
NGC4462    & SAB$_{\rm x}$(rs)a                                                               &    188 &   3.70 &    137 &   2.69 &   63.3 &  10.54 \\
NGC4569    & (R'L)SA$\underline{\rm B}_{\rm x}$(rs,x$_1$r)a                                   &    391 &  11.32 &    136 &   3.96 &   66.0 &  10.64 \\
NGC4586    & SA$\underline{\rm B}_{\rm x}$(s,nd)0/a sp                                        &    218 &   6.32 &     90 &   2.62 &   69.8 &  10.13 \\
NGC4725    & (R')SAB$_{\rm x}$(r,nb)a                                                         &    366 &  10.90$^*$ &    162 &   4.84 &   46.8 &  10.88 \\
NGC4818    & (RL)SA$\underline{\rm B}_{\rm xa}$(s)0${\circ}$                                  &    342 &   7.52 &    141 &   3.10 &   67.2 &  10.48 \\
NGC4845    & (R'L)SAB$_{\rm x}$(r'l,nd)0/a                                                    &    262 &  10.59 &    135 &   5.44 &   75.1 &  10.55 \\
NGC4856    & (RL)SB0$^-$                                                                      &    277 &   5.75 &    172 &   3.58 &   68.7 &  10.74 \\
NGC4902    & SB($\underline{\rm r}$s,bl)a$\underline{\rm b}$                                  &    206 &   4.23 &    209 &   4.30 &   21.5 &  11.05 \\
NGC5005    & (R$_{2}$')SAB$_{\rm xa}$(rs)ab                                                   &    396 &   5.99 &    244 &   3.70 &   66.7 &  11.10 \\
NGC5297    & SAB$_{\rm x}$(s)bc sp                                                            &     78 &   2.57 &    126 &   4.16 &   73.6 &  10.60 \\
NGC5443    & (R'L)SAB$_{\rm x}$(rs)a sp                                                       &    139 &   5.08 &    119 &   4.36 &   68.0 &  10.49 \\
NGC5448    & (R$_1$L)SAB$_{\rm x}$($\underline{\rm r}$s)a                                     &    167 &   8.43 &    143 &   7.24 &   65.4 &  10.77 \\
NGC5689    & (R'L)SAB$_{\rm x}$(r'l,nd)$\underline{\rm 0}$/a                                  &    181 &   5.85 &    171 &   5.53 &   74.4 &  10.84 \\
NGC5757    & (R')SB(rs)$\underline{\rm a}$b                                                   &    248 &   4.65 &    177 &   3.32 &   32.7 &  10.74 \\
NGC5806    & (R'L)SAB(rs,nrl)ab                                                               &    228 &   4.20 &    147 &   2.72 &   56.5 &  10.59 \\
NGC5854    & (RL)SA$\underline{\rm B}_{\rm x}$(rl)0$^+$ sp                                    &    189 &   3.76 &    119 &   2.37 &   71.1 &  10.23 \\
NGC5864    & (R$\underline{\rm L}$)SB$_{\rm xa}$0$^+$ sp                                      &    139 &   2.48 &    120 &   2.14 &   71.6 &  10.35 \\
NGC5878    & SAB$_{\rm xa}$(rs)ab                                                             &    185 &   5.13 &    149 &   4.14 &   68.1 &  10.77 \\
NGC7140    & (R')SA$\underline{\rm B}_{\rm x}$(rs,nrl)a$\underline{\rm b}$                    &    161 &   7.39$^*$ &    126 &   5.80 &   49.8 &  10.70 \\
NGC7163    & SAB$_{\rm x}$(s)a                                                                &    206 &   5.46 &    101 &   2.67 &   57.5 &  10.13 \\
NGC7171    & SAB$_{\rm x}$(s)b                                                                &     87 &   2.82 &    127 &   4.13 &   57.5 &  10.58 \\
NGC7179    & SB$_{\rm xa}$($\underline{\rm r'}$l)0/a                                          &    100 &   2.48 &    135 &   3.33 &   58.9 &  10.60 \\
NGC7421    & (R$^{\prime}$)SB(rs,bl)ab                                                        &    132 &   2.93 &    102 &   2.26 &   27.9 &  10.15 \\
NGC7513    & (R'L)SB(rs)a                                                                     &    102 &   3.09 &    101 &   3.05 &   48.4 &  10.21 \\
NGC7531    & SAB$_{\rm x}$(r)a                                                                &    221 &   3.90 &    144 &   2.55 &   57.5 &  10.44 \\
PGC045650  & SA$\underline{\rm B}_{\rm a}$(s)ab                                               &    185 &   2.38 &    136 &   1.76 &   72.7 &  10.37 \\
\enddata
\label{table}
\tablecomments{The inner slope $dV_{cir}/dr$ and the maximum velocity
  $V_{max}$ are calculated from the S$^4$G 3.6 $\micron$ images, the
  stellar masses $M*$ are from \citep{munozmateos2015}, the disk scale
  length and inclination are from \cite{salo2015}. Normalized $dV_{cir}/dr$
  indicates scaling of the inner slope with $V_{max}/h_r$: the X-shape
  galaxies with $i<60^\circ$ and normalized $dV_{cir}/dr>8$ are marked with
  an asterisk.  The barlens classifications are from \cite{buta2015}
  and \cite{laurikainen2011}, and the X classifications from
  \cite{laurikainen2016b}.
}

\end{deluxetable}



\end{document}